\documentclass[twocolumn,showpacs]{revtex4}
\usepackage{amssymb}
\usepackage{amsmath}
\usepackage[dvips]{graphicx}

\begin{document}

\title{Extraordinary magnetoresistance of  organic semiconductors  : Hopping conductance via non-zero angular momentum orbitals}
\author{A. S. Alexandrov$^{\star,1,2}$, V. A. Dediu $^{3}$, V. V. Kabanov $^{2}$, R. R. da Silva $^1$ and Y. Kopelevich $^1$}
\affiliation{$^1$ Instituto de Fisica "Gleb Wataghin", Universidade Estadual de Campinas, UNICAMP 13083-970, Campinas, Sao Paulo, Brasil\\
{$^{2}$ Josef Stefan Institute,
1001 Ljubljana, Slovenia\\
$^{3}$ ISMN-CNR, Via Gobetti 101, 40129 Bologna, Italy}}

\begin{abstract}
  Highly-anisotropic  in-plane magneto-resistance (MR) in  graphite (HOPG) samples has been recently observed (Y. Kopelevich \textit{et al.}, arXiv:1202.5642) which is negative and linear in low fields in some current direction  while  it is  giant, super-linear and  positive in the perpendicular  direction.  In the framework of the   hopping conductance theory via non-zero angular momentum orbitals we link   extraordinary MRs in graphite and  in organic insulators (OMAR) observed in about the same  magnetic fields. The theory predicts quadratic negative MR (NMR) when there is a time-reversal symmetry (TRS), and linear NMR if TRS is broken. We argue that the observed   linear NMR could be a unique signature of the broken TRS both in graphite and organic compounds.  While some  local paramagnetic centers  are  responsible for the broken TRS in organic insulators, a large diamagnetism of our HOPG samples may involve a more intriguing scenario of  TRS breaking.
\end{abstract}


\maketitle

\section{Introduction}
Transverse resistance, $R(B)$, of  isotropic conductors with  Bloch electrons  increases with an applied magnetic filed so that $MR\equiv R(B)/R(0)-1$ is positive and  quadratic in B \cite{ref}). A large  positive MR  is also used to be a hallmark of the hopping conductance in doped insulators \cite{shklo}.   While the carrier s-wave function on a donor (or acceptor) is spherically  symmetric in the absence of the magnetic field, it becomes cigar-shaped squeezed in the transverse direction to the field \cite{yaf,sladek,elliot,hasegava,miko}.  This leads to a significant decrease in the overlap of the wave-function tails of two neighboring donors, and hence to a significant  increase of resistivity (positive MR), which is also quadratic in low magnetic fields.  

However there are  exceptions which do not show these canonical MRs.  For instance some semiconductors and semimetals (e. g. bismuth), where  open Fermi surfaces are unfeasible, show  positive but linear MR. One of the theoretical possibilities for such a phenomenon is a so-called "quantum magnetoresistance" in semimetals  having  pockets of the Fermi surface with a small or even zero effective mass (as the Dirac fermions in bithmuth, graphite and graphene), which might be in the ultra-quantum limit at rather low magnetic fields \cite{abr}. 

Also there is anomalous \textit{negative} MR (NMR) observed in some hopping systems, for instance in amorphous germanium and silicon. Originally it has been attributed  to magnetic-field dependence of spin-flip transitions between sites when some fraction of them has a frozen spin \cite{movag}, and/or  to an increase of the density of localised states due to the Zeeman energy shift, $\mu_B B$ \cite{kam}. This NMR is used to be small (much less than 1\%) even in relatively high magnetic fields of about 1 Tesla. There are other theoretical mechanisms of NMR, in particular the weak localization  gives NMR which is  often almost linear in a certain field range.   Such NMR
smoothly evolves from a sub-linear magnetic field dependence
at lower temperatures to super-linear field
dependence at higher temperatures.
A strong NMR exists in
 the classical two-dimensional
electron gas due to freely circling electrons, which are not
taken into account by the Boltzmann approach. It is
parabolic  rather than linear  at low fields \cite{dyakonov}. The parabolic orbital NMR has been also predicted by the gauge theory in two-dimensional strongly-correlated doped Mott insulators \cite{wiegmann}.

More
recently a linear NMR has been observed in the longitudinal c-axis inter-layer current in the normal state of
cuprate superconductors and in graphite in high magnetic fields assigned to bipolarons in the former case \cite{zav}
and to a growing population of the zero-energy Landau
level of quasi-two-dimensional Dirac fermions with the
increasing magnetic field in the latter case \cite{kopbra}. Also a
giant transverse NMR of nearly 100$\%$ was observed at
low temperatures, with over 50$\%$ remaining at room
temperature in graphene nanoribbons and attributed to
some delocalization effect under the perpendicular magnetic field \cite{bai}.

Puzzling magnetoresistance
effects in several different $\pi$-conjugated polymer and
small molecular thin film devices have been observed named organic magnetoresistance (OMAR) \cite{mermer}. OMAR
reaches 10$\%$ at fields on the order of only 10 mT, and can be
either positive or negative, depending on operating conditions.
The effect is independent of the sign and direction of
the magnetic field. The Zeeman energy does not account for the observed OMAR at ambient temperatures since it  is too small, $\mu_B B \approx
 10$ mK,
in the field of 10 mT.  The observation of OMAR in hole-only devices \cite{mermer,bob} indicates, on the other hand, that the effect is hardly compatible with an exciton-based mechanism.

 An alternative model involving
 spin-dependent bipolaron formation in deep potential wells has been proposed
as a possibility \cite{bob}.  In this model bipolarons  (i.e doubly occupied sites)  block single polaron
transport through bipolaron states causing positive MR, while
 an increase in polaron population at the expense of
bipolarons with increasing magnetic field might cause negative MR,
if the long-range Coulomb repulsion around each carrier is
sufficiently strong. However, one might  expect that this mechanism
should depend strongly on the carrier density, whereas OMAR is only
weakly dependent on current density \cite{mermer}.    Finding a convincing explanation of OMAR  is crucial for a better understanding of charge transport in organic semiconductors, actively used in   light-emitting diodes, photovoltaic
cells, field-effect transistors, and in  spintronics \cite{dediu,wohl}.

Recently an unusual orbital MR 
in HOPG  has been observed \cite{kop}. In some current direction it is  negative and linear  in low fields  with the crossover to the positive MR at higher fields, while in the perpendicular current direction MR is  giant, super-linear and   positive.
In this paper we compare  OMAR and graphite MR and propose an explanation of both MRs in the framework of   the  hopping magneto-conductance via non-zero angular momentum orbitals \cite{aleomr}.

\section{Hopping magnetoconductance via non-zero angular momentum orbitals}
In   organic and inorganic doped insulators  lattice defects such as
vacancies,  interstitials, and excess neutral atoms or ions often localise
carriers in  finite momentum states rather than in the zero-momentum
s-state \cite{physics}.  Also the conventional nonmagnetic donors
and acceptors have nonzero orbital momentum states along
with s-states, which are accessible for the hopping
conduction at elevated temperature. Here we briefly outline the   theory of hopping MR
via non-zero angular momentum orbits \cite{aleomr}.

In the hopping regime with localized carriers MR is
caused by a strong magnetic field dependence of the exponential
asymptotics of  bound state wave functions at a remote distance
from a donor (or an acceptor). The asymptotics is found using the
 Green function, $G(\textbf{r},\textbf{r}^\prime;E)$, (GF) of free (or  Bloch) electrons in the magnetic field with a negative binding energy $E$ \cite{shklo}. 
 The three-dimensional GF, $G_{3D}(\textbf{r},\textbf{r}^\prime;E)$  is readily calculated using  the Fourier transformation of the two-dimensional $G_{2D}\left
(\overrightarrow{\rho},\overrightarrow{\rho}^\prime;E\right)$ \cite{dod} ,
\begin{equation}
G_{3D}(\textbf{r},\textbf{r}^\prime;E)= {1\over{2\pi
\hbar}}\int_{-\infty}^{\infty} dp e^{ip (z^\prime -z)}
G_{2D}\left
(\overrightarrow{\rho},\overrightarrow{\rho}^\prime;E-{p^2\over{2m_b}}\right), \label{GF}
\end{equation}
\begin{eqnarray}
&&G_{2D}(\overrightarrow{\rho},\overrightarrow{\rho}^\prime;E)={m_b\over
{2\pi \hbar ^2}}\exp\left[i {\rho \rho'
\sin(\phi'-\phi)\over{2l^2}}\right] \times \cr &&
e^{-(\overrightarrow{\rho}-\overrightarrow{\rho}^\prime)^2/4l^2}
\Gamma(a)U[a,1,(\overrightarrow{\rho}-\overrightarrow{\rho}^\prime)^2/2l^2],
\label{gf}
\end{eqnarray}
where  $\phi$ and $\phi^\prime$ are azimuth angles of
$\overrightarrow{\rho}$ and $\overrightarrow{\rho}^\prime$
respectively, $l=(\hbar/eB)^{1/2}$ is  the magnetic length,
$\Gamma(a)$ is the gamma-function,  $U(a,b,z)$ is the Tricomi's
confluent hypergeometric function well-behaved at infinity,
$z\rightarrow \infty$, for negative $E$,  and
$a=1/2-(E\mp \mu_BB)/\hbar \omega_c$ ($\mp$ corresponds to
spin up/down, respectively). Neglecting a small
diamagnetic correction (quadratic in $B$)
 yields 
 \begin{equation}
 E=-\epsilon_0 + \hbar \omega_c m/2 \pm \mu_BB,
 \end{equation}
  where $\omega_c=eB/m_b$ ($m_b$ is the band mass), $m=0, \pm 1,
\pm 2,...$ is the magnetic quantum number of the localised state
so that
 \begin{equation}
a= {1\over{b}} +{1-m\over{2}}.
\end{equation}
Hereafter we measure the magnetic field, $b=B/B_0$,  in units of $B_0=\hbar\kappa^2/2e$, where  $\kappa=(2m_b
\epsilon_0)^{1/2}/\hbar$ is the inverse localisation length of the
zero-field  state with the ionisation energy $\epsilon_0$.

Now using an integral representation of the confluent
hypergeometric function \cite{abram}   one obtains  after integrating over $p$ in Eq.(\ref{GF}) 
\begin{eqnarray}
&&G_{3D}(\textbf{r},0;E)={m_b \over {(2\pi)^{3/2}
\hbar^2}l}\int_{0}^{\infty} dx {e^{mx}\over{\sqrt{x} \sinh (x/2)}}
\times \cr &&\exp\left\{-\left [{(\kappa \rho)^2 b\over{8}}
+{x\over{b}}+{(\kappa z)^2 b\over{4x}}+{(\kappa \rho)^2 b\over{4
(e^x-1)}}\right]\right\}. \label{gf3D}
\end{eqnarray}
Expanding the exponent in the square brackets in Eq.(\ref{gf3D}) up
to the third power in $x$ and performing the integration by the
saddle-point method we  obtain the asymptotics of the 3D impurity wave
function, $\psi(\textbf{r},B) \propto G_{3D}(\textbf{r},0;E)$ at $1\ll \kappa r \lesssim 1/b$ as \cite{aleomr}
\begin{equation}
\psi_m(\textbf{r},B)\propto \psi_m(\textbf{r},0) {\kappa r
b/2\over{\sinh(\kappa r b/2)}}\exp \left[{m \kappa r
b\over{2}}-{\kappa^3 \rho^2 r b^2\over{96}}\right]\label{3Dpsi}
\end{equation}
with $r^2=\rho^2+z^2$. 

For the s-wave bound state with $m=0$ this is
the textbook asymptotics \cite{miko,shklo} accounting for the
conventional positive MR quadratic in small $B$ . On the contrary,
for orbitals with nonzero orbital momentum the wave function,
Eq.(\ref{3Dpsi}), is \textit{linear} in small $B$. 
Remarkably a localized state with a positive $m$ expands in the magnetic field  due to the 
magnetic lowering of its ionization energy by $\hbar \omega_c m/2$, while  states with  negative $m$s shrink. The linear term
 dominates in a wide range of realistic
impurity densities for any nonzero  $m$ because of the small numerical factor ($1/96$) in the quadratic term in the exponent  of Eq.(\ref{3Dpsi}).

\begin{figure}
\begin{center}
\includegraphics[angle=-0,width=0.50\textwidth]{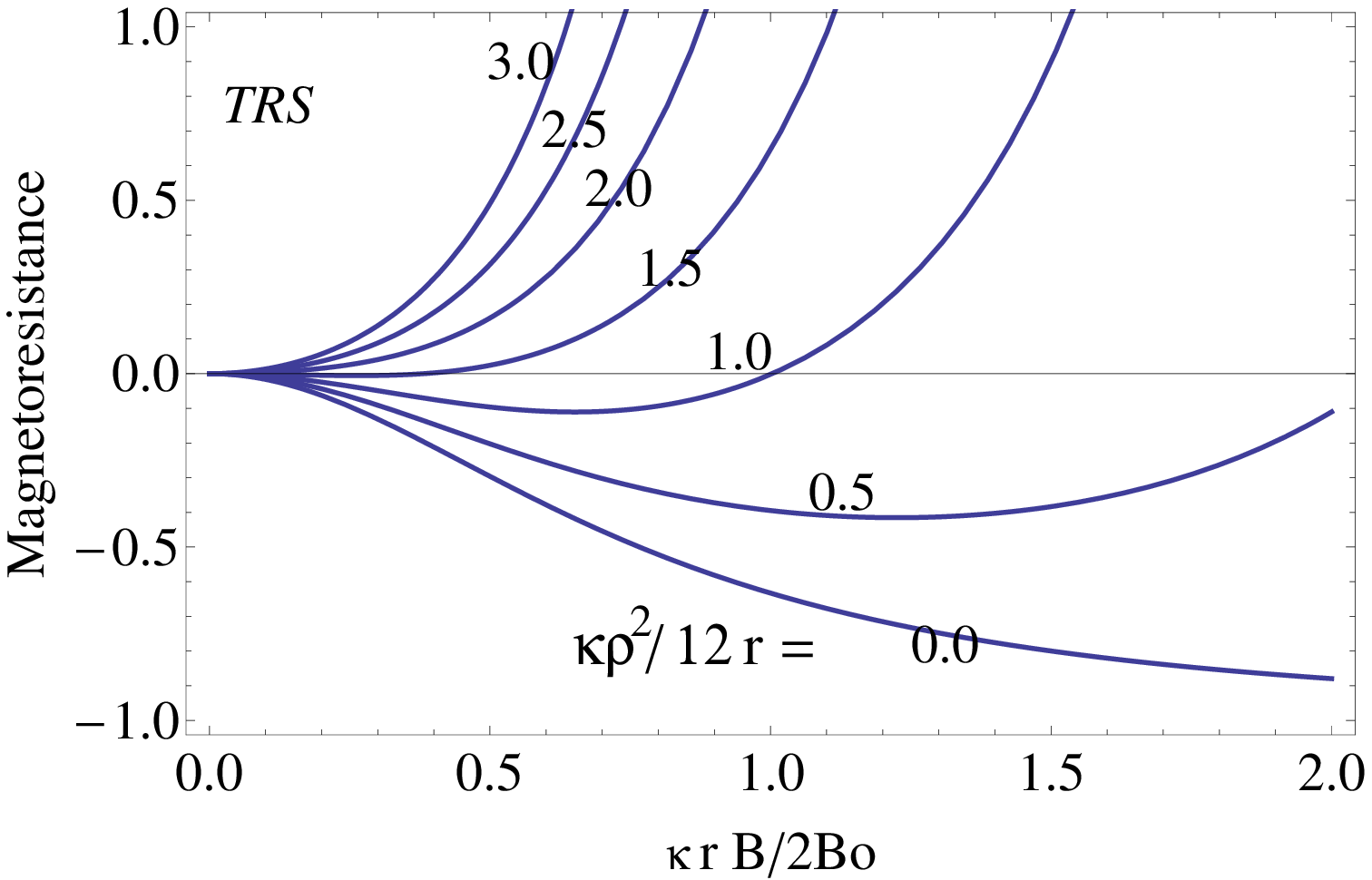}
\includegraphics[angle=-0,width=0.50\textwidth]{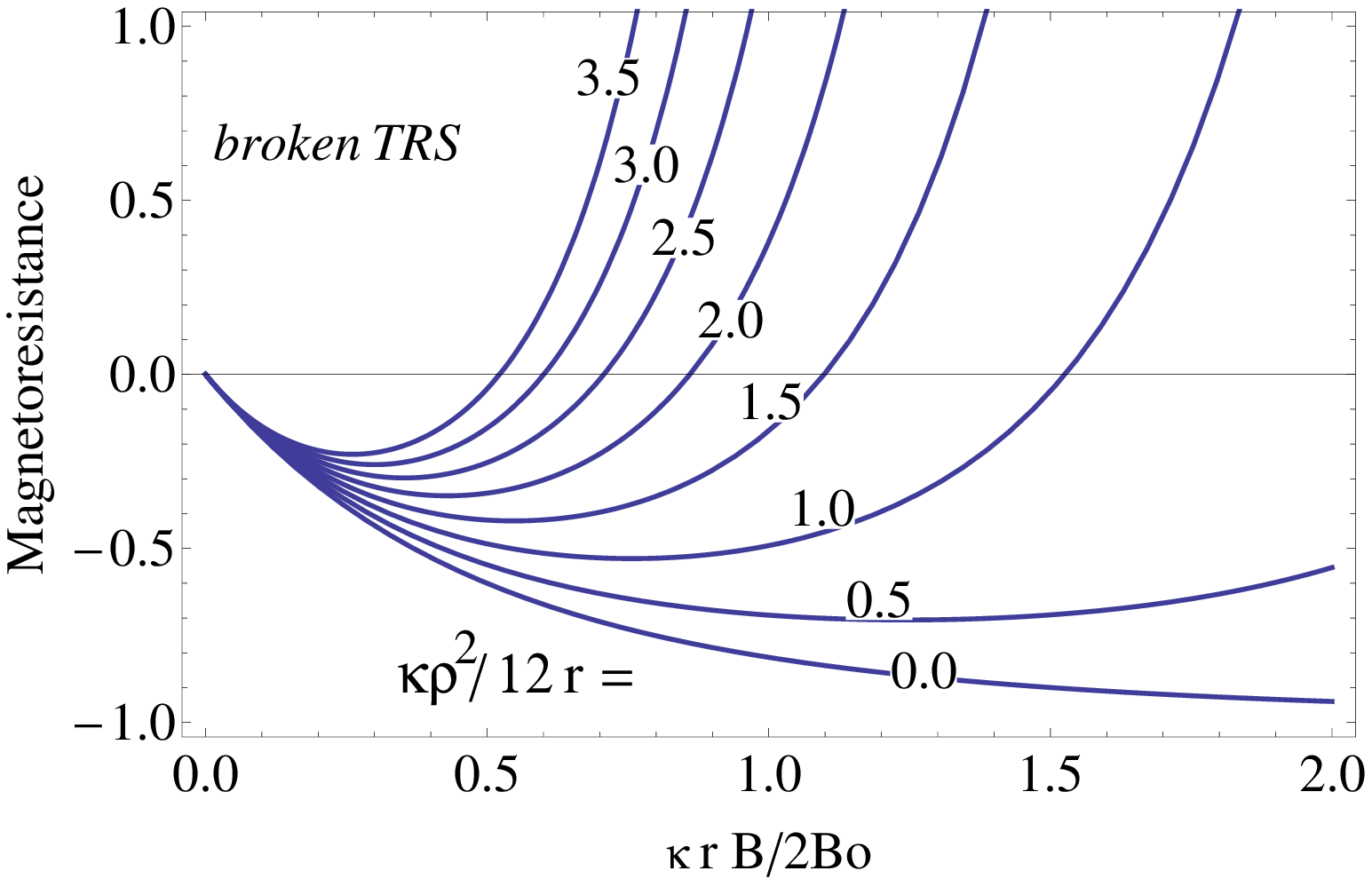}
\vskip -0.5mm \caption{Upper panel: Magnetoresistance to hopping via $p$- orbitals  in TRS systems versus the magnetic field. Lower panel: The same MR in systems with the broken TRS. }  \label{TRS}
\end{center}
\end{figure}

The hopping integral is proportional to 
$\psi_m(\textbf{r},B)$, where $r$ is the distance between two hopping
sites, which is assumed to be large  compared with the
localisation length. The hopping conductance is proportional to the
hopping integral squared.
If there is no time-reversal symmetry breaking,   the states with
the opposite direction of the orbital angular momentum, $m$ and
$-m$, are degenerate, so that the linear term in the conductivity,
$\sigma=\sigma_m +\sigma_{-m}$ cancels, and
\begin{equation}
\sigma (B)=\sigma(0)\left[{\kappa r b/2\over{\sinh(\kappa r
b/2)}}\right]^2 \cosh (m \kappa r b)e^{-\kappa^3 \rho^2 r b^2/48}.
\label{sigma}
\end{equation}
In this case  the hopping conductivity, $\sigma(B)$ first
\emph{increases} quadratically with the magnetic field (parabolic NMR) and
only then decreases with $B$ (positive MR), if $\kappa
\rho^2/r < 24m^2-4$ as illustrated in  Fig.\ref{TRS} (upper panel) representing $MR=\sigma(0)/\sigma(B)-1$ for  p-states ($m=\pm 1$). 

Ions that carry a magnetic
moment break the time-reversal symmetry and split $m$ and $-m$
states. Such zero-field splitting (ZFS) gives preference to 
hopping via orbitals with a lower ionisation energy (positive $m$).
 Hence in a
ferromagnet with the \textit{frozen} magnetisation MR
for hopping via nonzero momentum orbitals is highly anisotropic
changing from linear and negative in the field applied parallel to
the magnetisation to linear but positive in the opposite field. On the other hand if the global or local internal magnetisation  rotates with the external magnetic field (superparamagnetism) magnetoresistance is negative and linear in small $\textbf{B}$ no matter what  the direction of the external field is. When
the  splitting due to an internal (exchange) field is comparable or larger than  
the temperature, the hopping MR is found as
\begin{equation}
MRh=-1+ \left[{\sinh(\kappa r
b/2)\over{\kappa r b/2}}\right]^2 \exp\left[\kappa^3 \rho^2 r {b^2\over{48 }}-|m|\kappa r {b}\right].
\label{NMR}
\end{equation}
MRh in systems with the broken TRS, Eq.(\ref{NMR}), shown in Fig.\ref{TRS} (lower panel), is quite distinguishable  from MRh in TRS systems, Fig.\ref{TRS}(upper panel).
\begin{figure}
\begin{center}
\includegraphics[angle=-0,width=0.50\textwidth]{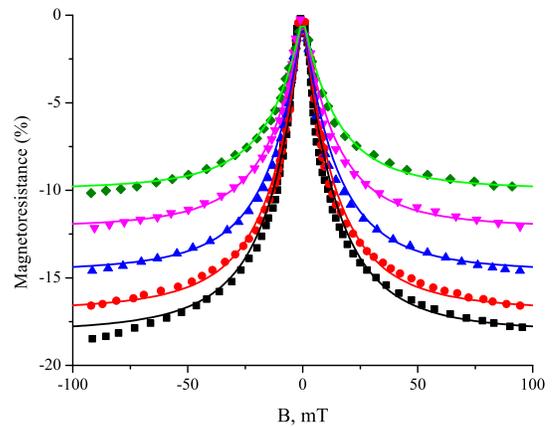}
\vskip -0.5mm \caption{(Color online) Magnetoresistance measured at room
temperature in  
$ITO/PEDOT /Alq3/Ca$ device at  bias voltages $12,13,14,15,17$ V  (symbols from  bottom to top, respectively,
Ref.\cite{mermer}) described by Eq.(\ref{NMR2}) (solid lines) with 
$B_h=30$ mT, $m=1$ and the resistance ratio $R$ shown in Fig.\ref{R}. }  \label{omarfit}
\end{center}
\end{figure}

\section{OMAR}
 Eq.(\ref{NMR}) allows  us to address  puzzling
experimental observations of negative NMR in a
number of  organic materials \cite{mermer,bob,ngu}. There is
experimental  evidence for paramagnetic centers and ZFS in polymers,
in particular  in Alq3 \cite{para}. More recently  dilute magnetic
impurities and
 magnetic domains have been observed in some $\pi$-conjugated polymers \cite{FM}. Hence as suggested in Ref.\cite{aleomr}, if TRS is broken by such magnetic centers, the low-field  magnetoresistance is dominated by the linear NMR via non-zero angular momentum orbitals, Eq.(\ref{NMR}). Here we extend our original description \cite{aleomr}  of  OMAR to the whole range of magnetic and electric  fields used in the experiments. 
 
The device fabrication \cite{mermer,bob,ngu} started with glass substrates coated with a metallic layer, indium-tin-oxide ITO or the conducting polymer PEDOT. The semiconducting polymer (e. g. Alq3) layer was thermally evaporated onto the bottom electrode, yielding an organic-semiconductor layer thickness of about 100 nm. The cathode, either Ca with an Al capping layer, Al, or Au was then deposited on top. 

The resistance of the device is  the sum of the hopping resistance, $R_{h}$, of the organic semiconductor layer probably
including the interface,  and of other  layers and contacts,  $R_{m}$. Hence, the magnetoresistance  is expressed via metallic/contact   $MRm=R_m(B)/R_m(0)-1$ and  the hopping magnetoresistance, $MRh$ of Alq3 as
 \begin{equation}
 MR= {MRm\over{1+1/R}}+{MRh\over{1+R}}, \label{MRomar}
 \end{equation}
 where $R=R_m(0)/R_h(0)$ is the ratio of the zero-field  resistance of other layers and contacts to the zero-field organic-semiconductor resistance. Reducing the number of fitting parameters we further assume that $MRm$ is small, $MRm \ll MRh$, and the first term in Eq.(\ref{MRomar}) can be neglected. Also as outlined above the quadratic term in the exponent of Eq.(\ref{NMR}) can be dropped in the relevant magnetic fields which yields
\begin{equation}
MR\approx {(2B_h\sinh(B/2B_h)/B)^2 \exp(-|m|B/B_h)-1\over{1+R}},
\label{NMR2}
\end{equation}
where $B_h=B_0/\kappa r$ is the characteristic scaling field.
 
\begin{figure}
\begin{center}
\includegraphics[angle=-0,width=0.40\textwidth]{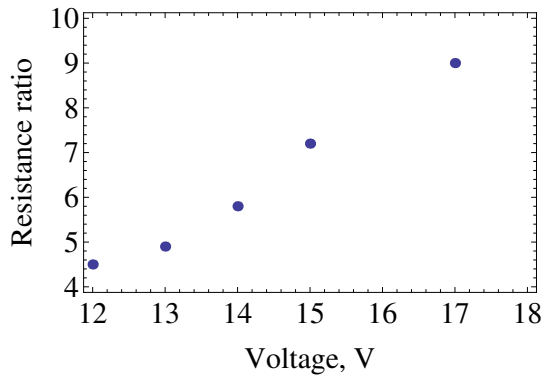}
\vskip -0.5mm \caption{Ratio of the zero-field  resistance of other layers and contacts to the zero-field organic-semiconductor resistance. }  \label{R}
\end{center}
\end{figure}

As shown in Fig.\ref{omarfit}  Eq.(\ref{NMR2}) with  $R$ as a single fitting parameter describes remarkably well the experimental NMR for all magnetic and electric fields used in the experiment.  The resistance ratio $R$ increases with  the voltage, $V$, on the device, Fig.\ref{R}, presumably due to  a drop of the zero-magnetic field $R_h(0)$ with $V$.
 
\section{Graphite MR}
Ref.\cite{kop} observed unusual MR  in  commercially available HOPG samples with different mosaicity and the room temperature out-of-plane/basal plane zero-field resistivity ratio $\rho_c/\rho_{ab}$ as high as  $10^5$.  The magnetic field was applied parallel to the hexagonal c-axis ($B \parallel c$), and  the in-plane $\rho_{ab}(B,T)$ was measured placing  silver pasted electrodes  on the sample surface.  All resistance measurements were performed in the Ohmic regime (more sample details and complementary magnetization measurements are found in Ref.\cite{kop}).
\begin{figure}
\begin{center}
\includegraphics[angle=-0,width=0.50\textwidth]{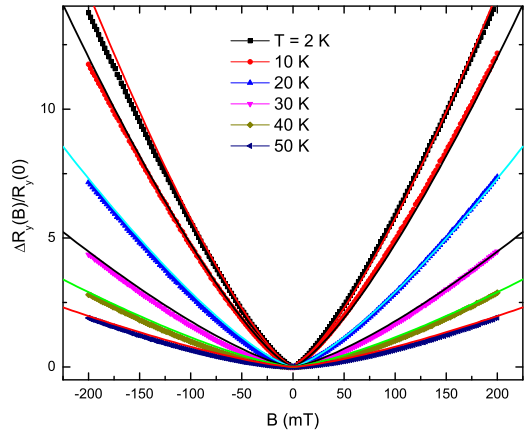}
\vskip -0.5mm \caption{(Color online) Giant in-plane  magnetoresistance of HOPG in the metallic (Y) direction at different temperatures (symbols \cite{kop}) fitted with $(B/B_{in})^{4/3}$ (lines).}  \label{mry}
\end{center}
\end{figure}

The graphite samples described in Ref.\cite{kop} show qualitatively different temperature dependence of the in-plane zero-field resistance for the current in two perpendicular in-plane directions. In one direction (called here  Y) the resistance is metallic-like
while in the other direction (X) it is insulating-like in a wide temperature interval from  2K to the room temperature. Since a sufficiently strong magnetic field of about 1 Tesla or more makes graphene planes electrically isotropic the anisotropy is of electronic origin.   The in-plane electrical anisotropy is most probably associated with an inhomogeneous carrier-density distribution, such that well doped metallic clusters are partially overlapped in Y-direction while they are separated by poorly doped insulating regions in X-direction. The in-plane anisotropy has been observed only in most anisotropic ($\rho_c/\rho_b \approx 10^5$)  graphite samples pointing  to its quasi-2D origin. 
Our measured Kish and natural graphite crystals possess much lower resistivity ratio ($\rho_c/\rho_b \approx 10^2$ -$ 10^3$) with no such effect  observed. No in-plane anisotropy has been  observed in  HOPG samples with $\rho_c/\rho_b < 10^4$ either. It brings us to a conclusion that the electrical anisotropy  is closely  related to the reduced electron  dimensionality in quasi-2D graphite providing a quasi-1D percolation in the  planes.

\begin{figure}
\begin{center}
\includegraphics[angle=-0,width=0.40\textwidth]{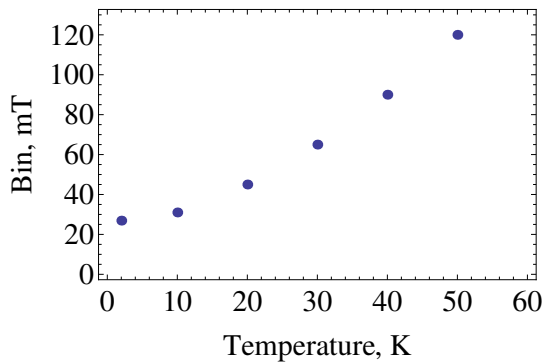}
\vskip -0.5mm \caption{ Characteristic field $B_{in}$ of  metallic clusters in HOPG as a function of temperature.   } \label{Bin}
\end{center}
\end{figure}

When a relatively weak magnetic field is applied perpendicular to the planes, the MR along the metallic Y direction appears huge and positive, Fig.\ref{mry}. A giant positive MR is naturally expected in doped graphite with  virtually massless Dirac fermions \cite{koplyk} since the parameter $\beta=\omega_c \tau$ becomes large already in the mT-region of the field (here  $\tau$ is the scattering time). However, it is neither quadratic as  in the Boltzmann theory, nor linear in B as in the "quantum magnetoresistance" \cite{abr}.  It has been  suggested \cite{kop} that inhomogeneities are responsible for the strong departure from these  regimes.
They lead to a radical rearrangement of the current flow pattern  changing the magnetic field dependence of the transverse conductivity.  Importantly, when  $\beta > 1$ , then  even  relatively small inhomogeneities in the carrier density   lead to  the MR  proportional to $B^{4/3}$ \cite{dykhne}. In fact,  $MRy=(B/B_{in})^{4/3}$ perfectly fits  the observed magnetic field dependence of MR in the metallic current direction, Fig.(\ref{mry}) with a single scaling parameter $B_{in}$ depending on  fluctuations in the carrier density \cite{dykhne} and temperature, Fig.\ref{Bin}. The temperature dependence of $B_{in}(T)$ is reminicent of the temperature dependence of metallic resistivity, as it should be since $B_{in} \propto 1/\tau$ \cite{dykhne}.

\begin{figure}
\begin{center}
\includegraphics[angle=-0,width=0.38\textwidth]{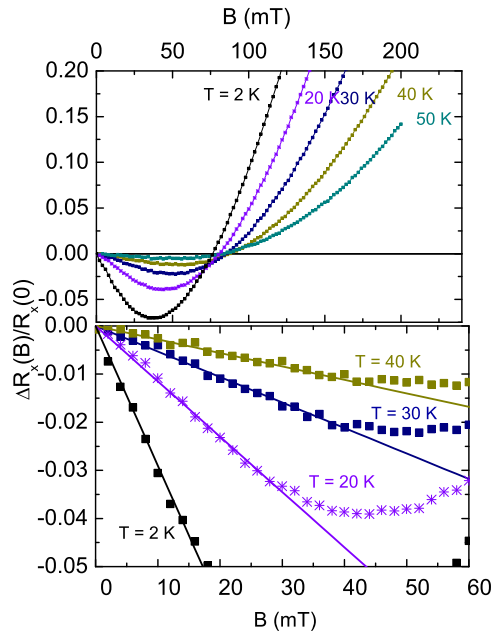}
\vskip -0.5mm \caption{(Color online)  Upper panel: Magnetoresistance in the insulating (X) direction (symbols). Lower panel: Low-field   negative linear MR fitted  as $MRx= - B/B_s$   with $B_s=(1+R)B_h/|m|$, shown in Fig.\ref{Bs} at different temperatures,  indicating the time-reversal symmetry breaking.} \label{mrx}
\end{center}
\end{figure}

 The MR in the insulating direction is also anomalous, Fig.\ref{mrx}. It is  linear and negative in very low fields and positive and superlinear in higher fields above the crossover point. The insulating-like temperature dependence of the zero-field resistance \cite{kop} in this direction supports the view that the metallic clusters are virtually nonoverlaped along X, and  the resistance is  the sum of the hopping resistance, $R_{h}$ of the insulating regions with low doping, and the metallic-like resistance $R_{m}$. Hence, the magnetoresistance in X direction is expressed via metallic  MRy and the hopping magnetoresistance, MRh of insulating layers as in Eq.(\ref{MRomar}),
 \begin{equation}
 MRx= {MRy\over{1+1/R}}+{MRh\over{1+R}}, \label{mrxtheory}
 \end{equation}
 where $R=R_m(0)/R_h(0)$ is the ratio of the zero-field metallic resistance to the zero-field insulating resistance. This ratio ranges from about $0.1$ at $2K$ to $1$ at room temperature \cite{kop}.
 
 However, different from OMAR Eq.(\ref{MRomar}) the first term in Eq.(\ref{mrxtheory}) is signifincat and responsible for the upturn of $MRx$ from negative to positive at some magnetic field about 70 mT, Fig.(\ref{mrx}) as described in Ref.\cite{kop}. Here we analyse in more detail the low-field asymptotics of $MRx$. If TRS is broken in HOPG then according to Eq.(\ref{NMR2}) this asymptotics is linear in B because the first term in Eq.(\ref{mrxtheory}) is superlinear,
 \begin{equation}
 MRx\approx -{|m|B\over{(1+R)B_h}}
 \end{equation}
 for $B \rightarrow 0$.

\begin{figure}
\begin{center}
\includegraphics[angle=-0,width=0.49\textwidth]{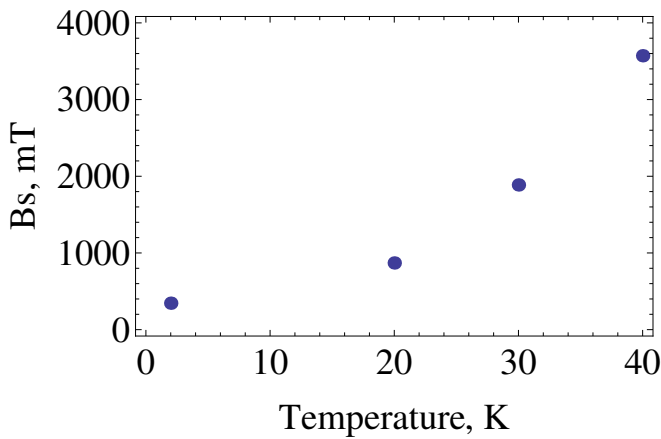}
\caption{Inverse slope $B_s$ of the linear NMR in HOPG versus temperature.} \label{Bs}
\end{center}
\end{figure}

With increasing temperature the resistance ratio $R$ increases and deeper localized states with a higher ionization energy become accessible  for the hopping conductance, so that the slope  $1/Bs$ of the linear negative MR drops, as observed, Figs.(\ref{mrx},\ref{Bs}).

\section{Conclusions}
In conclusion, we described the unusual magnetoresistance of organic insulators (OMAR) \cite{mermer,bob,ngu}  and  the highly-anisotropic  in-plane magnetoresistance 
of quasi-two dimensional HOPG graphite \cite{kop}  using the same theory of hopping magneto-conductance via non-zero angular momentum orbitals \cite{aleomr}.  While the theory  with one or two scaling parameters provides accurate agreement with OMAR and the HOPG data, there are other theoretical mechanisms of NMR outlined in the introduction and beyond. Most of them explain NMR as some spin-correlation effects.
Importantly, there is no NMR and virtually no MR in the same range of the magnetic field \emph{parallel} to the planes, which rules out a spin origin of the observed NMR in our quasi-2D HOPG samples. Also there are no quantum magnetic oscillations at  high temperatures, where  unusual MRs are still observed, Figs.(\ref{mry},\ref{mrx}), so they are not related to the Landau quantization.  In contrast to a number of spin correlation  and weak localisation scenarios we observed the perfectly linear NMR at very low $B$ in a wide temperature range, Fig.(\ref{mrx}), which according to our theory is a clear signature of the time-reversal symmetry breaking. 

The origin of the broken TRS in organic compounds and  in graphite could be basically  different. Instead of the (super)paramagnetism in polymers \cite{para,FM}
the graphite samples of Ref. \cite{kop} show a large diamagnetism. Naturally, some  local paramagnetic centers responsible for the broken TRS in organic semiconductors could be also found  in  graphite, with their magnetic response  overwhelmed  by the large diamagnetism of metallic clusters.  However the observed large diamagnetism could suggest a more intriguing mechanism of  TRS breaking, such as    superconducting clusters \cite{kopsp} with an unconventional (chiral)  order parameter \cite{don,chub}.  There is a kink in the field dependence of the diamagnetic magnetization of the HOPG samples at $B_k \approx 0.2$ T, \cite{kop}, resembling the behavior of  type-II superconductors in magnetic fields  exceeding the lower critical field. Supporting this possibility the electrically inhomogeneous samples are becoming  homogeneous insulators in sufficiently high magnetic fields, which could suppress the superconductivity. Also in some   2D lattices there is the broken time-reversal
symmetry (i.e. some internal magnetic ordering) in the normal state with a spontaneous quantum Hall effect but without any net magnetic
flux  at high temperatures \cite{haldane}. At lower temperatures a pair of spontaneously
generated current loops in adjacent graphene layers, having odd-parity with respect to the two layers, could also break TRS \cite{varma}
  
  More generally our findings point to an inhomogeneous doping and a semiconducting gap in graphite.

\section*{Acknowledgemets}
This work has been supported by FAPESP, CNPq, CAPES, ROBOCON, INCT NAMITEC, the European Union Framework Programme 7
(NMP3-SL-2011-263104-HINTS), and
 by the UNICAMP visiting professorship programme.

\end{document}